\documentclass[aps,prl,twocolumn,showpacs,superscriptaddress,groupedaddress]{revtex4}  
\usepackage{graphicx}  
\usepackage{dcolumn} 
\usepackage{bm}     
\usepackage{amssymb}  

\usepackage{amsmath}
\begin{document}

\title{Stability of Branched Flow from a Quantum Point Contact}

\author{Bo Liu}
\email{bliu@physics.harvard.edu}
 \affiliation{Department of Physics, Harvard University, Cambridge, Massachusetts 02138, USA}

\author{Eric J. Heller} \affiliation{Department of Physics, Harvard University, Cambridge, Massachusetts 02138, USA}
\affiliation{Department of Chemistry and Chemical Biology, Harvard University, Cambridge, Massachusetts 02138, USA}
\date{\today}

\begin{abstract}
In classically chaotic systems, small differences in initial conditions are exponentially magnified over time. However, it was observed experimentally that the (necessarily quantum) ``branched flow'' pattern of electron flux from a quantum point contact (QPC) traveling over a  random background potential in two-dimensional electron gases(2DEGs) remains substantially invariant to  large changes in initial conditions. Since such a potential is classically chaotic and unstable to changes in initial conditions, it was conjectured that the origin of the observed stability  is  purely quantum mechanical, with no classical analog. In this paper, we show that the observed stability is a result of the physics of the QPC and the nature of the experiment. We  show that the same stability can indeed be reproduced classically, or quantum mechanically. In addition, we explore the stability of the branched flow with regards to changes in the eigenmodes of quantum point contact.
\end{abstract}

\pacs{ 73.23.-b, 73.63.-b, 05.45.Mt}
\maketitle

Branching is a universal phenomenon of wave propagation in a weakly correlated random medium. It is observed in 2DEGs with wavelength on the scale of nanometers\cite{a2,a3}, in quasi-two-dimensional resonator with microwave\cite{a12} and used to study sound propagation in oceans with megameter length scales\cite{a10}. It has significant influence on electron transport in 2DEGs\cite{a14,a13} and is found to be implicated in the formation of freak waves in oceans\cite{a9}.  In all these studies, classical trajectory simulations show closely similar branch formation.
\begin{figure*}
\includegraphics[width=14cm]{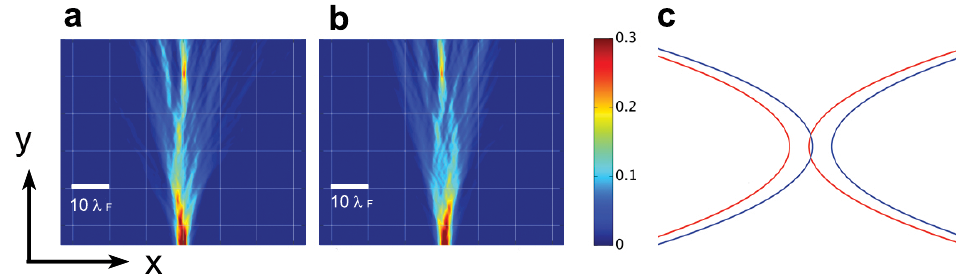}
\caption{(Color Online) Quantum simulations of the total flux in the y direction that passes through a given point.  ({\bf a}) plots the case where the QPC is not shifted and ({\bf b}) shows the case where the QPC is shifted by $\lambda_{F}$ to the right . The white reference grid(10$\lambda_{F}$) denotes the same location in both images. The color axis shows the normalized density of flux per wavelength. Both images start at y=$y_{0}+15\lambda_{F}$. More information about these images can be found in the supplementary material. ({\bf c}) A schematic plot of the QPCs used in the simulations(Red in ({\bf a}) and blue in ({\bf b})).}
\label{qflux}\end{figure*}

However, the classical interpretation was challenged by a recent experiment on 2DEGs\cite{a1}, where it was observed that the (necessarily quantum) branched flow pattern even far away showed stability of the branches against the changes in the QPC. This stability was conjectured to be of quantum origin\cite{a1}. To our best knowledge, no insights into this stability have been provided since, and it remains a puzzle in the literature.  In this paper, we provide an explanation for the observed stability. Moreover, we provide numerical simulations to show that it can indeed be reproduced by classical trajectories.

To proceed, we need to recount  what was done in the experiment\cite{a1}. To create a large change in initial conditions, the QPC was shifted by about one correlation length of the underlying random potential, which is also roughly the width of the QPC. Classically, a one correlation length shift is indeed very significant for the chaotic dynamics, making the trajectories very different, as seen in the classical simulations of reference\cite{a1}. If one launches two separate quantum wavepackets\cite{a4} through  QPCs differing by this amount, the coherent overlap between the two initial wavepackets is estimated at less than five percent. However, in the experiment it is nonetheless observed that some branches remain at almost exactly the same locations seventy correlation lengths away from the injection points, with the only observed difference being the relative strength of each branch. This lack of sensitivity even at long range was termed ``the unexpected features of the branched flow''. 

The present paper gives an explanation to the observed stability, by taking into account the effect of the change of QPCs. As a Gedanken experiment, consider a pair of side-by-side QPCs differing by a shift. This could not be                                                                                                                                                                                                                                                                                                                                                                                               in the experiment, which had only one QPC, which however was able to be shifted relative to the rest of the device and the branched flow imaged again. In the Gedanken experiment, suppose we put wavepacket A though one QPC and wavepacket B through the other. Can the coherent overlap between  the initially nonoverlapping A and B  wavepackets  increase over time and distance from the QPC's? The answer is of course no, both classically (considered as overlap in phase space) and quantally.  It is elementary to show that the coherent overlap must remain the same over time if the wavepackets are propagated under the same Hamiltonian. This is true whether or not disorder is present. 

However, in the experiment as performed, \textit{the Hamiltonian of a single QPC and the Hamiltonian with the  QPC shifted over are not the same.} Therefore, no theorem constrains the evolution of the coherent overlap between the two different initial wavepackets. As it will become clear, this is exactly what leads to the observed stability. 

In order to show this, we first consider the ideal case where the QPC is perfectly adiabatic and provide an analytical solution of the coherent overlap between the two wavepackets launched from \textit{two different} QPCs as a function of time. We show the correlation reaches almost one at sufficiently large distance even if the initial coherent overlap is negligible.  We then choose a more realistic QPC potential and also add smooth disorder of the type causing the branching into the system.  The coherent overlap in this case still  reaches $85\%$. Finally, we show that the same mechanism works for classical trajectories. For the classical case, we calculate an overlap of 79$\%$ in the phase space. Both results prove that the stability in the experiment is due to the nature of the experimental QPC shift. In the last part of the paper, we also make a prediction on the stability of the branched flow when the second mode in the QPC is open.

For a QPC with harmonic confinement, the Hamiltonian is
\begin{equation} 
\begin{aligned}
H_{o}=\frac{\vec{p}^{2}}{2m}+\frac{1}{2}m\omega^2(y)x^2
\end{aligned}
\label{H} \end{equation}
where $\omega(y)$ is a slowly varying function of y and decreases monotonically as the QPC opens up.   According to the approach developed in \cite{a4}, we can reproduce the experimental results by propagating an initial wavepacket of the following form through the system.
\begin{equation} 
\begin{aligned}
\Psi_{o}(x,y,0)&=\int dE \ e^{i\varphi (E,y_{0})}\sqrt{-\frac{m}{2\pi\hbar^2}\frac{\partial f_{T}(E,E_F)}{\partial E}}\Psi_{1}(x,y,E)
\end{aligned}
\label{iwave} \end{equation}
where $\Psi_{1}(x,y,E)$ is the scattering eigenstate at energy E , $\varphi (E,y_{0})$ is chosen so that it is a compact wavepacket centered at $y=y_{0}$, and $f_{T}(E,E_F)$ is the Fermi distribution with temperature T and Fermi energy $E_F$. This is a so-called ``thermal wavepacket'' at temperature T.  For one mode open in the QPC, the thermal wavepacket gives the correct thermally averaged conductance, by propagating  the wavepacket through the scattering region and counting the total flux that passes through a given point. More information about this method can be found in both reference \cite{a4} and the supplementary material.

First we  consider a perfect QPC and assume that disorder is absent. Numerical results including both disorder and an imperfect QPC follow. For a perfectly adiabatic QPC satisfying $\frac{\omega'(y)}{\omega^{2}(y)}\ll\frac{m}{\hbar k_{F}}$ and $\frac{\omega''(y)}{\omega^{2}(y)}\ll\frac{m}{\hbar}$, where $k_F$ is the Fermi wavevector,  $\Psi_{1}(x,y,E)$ can be approximated by
\begin{equation} 
\begin{aligned}
\Psi_{1}(x,y,E)=\frac{A(y_{0})}{\sqrt{\hbar k(y,E)\sqrt{\pi}\sigma(y)} }e^{i\int_{y_{0}}^{y}k(y',E)dy'-\frac{x^{2}}{2\sigma(y)^{2}}}
\end{aligned}
\label{eigenchannel} \end{equation}
where $\sigma(y)=\sqrt{\frac{\hbar}{m\omega(y)}}$, $\frac{\hbar^2 k^2(y,E)}{2m}+\frac{1}{2}\hbar w(y)=E$ and $A(y_0)$ is the normalizing constant. 

 The effect of shifting the QPC is incorporated in the initial wavepacket as
\begin{equation} 
\begin{aligned}
\Psi_{s}(x,y,0)=\hat{L}(x_{0})\Psi_{o}(x,y,0)
\end{aligned}
 \end{equation}
where $\hat{L}(x_{0})=e^{-ix_{0} \hat{p}_{x}/\hbar}$ is the translation operator, $x_0$ is the displacement of the QPC and $\hat{p}_{x}$ is the momentum operator in the x direction.

The two wavepackets evolve under the influence of their respective QPC and the coherent overlap between them at a later time t is given by
\begin{equation} 
\begin{aligned}
C_{o,s}(t)=&\left | \int dxdy\ \Psi^{*}_{o}(x,y,t)\Psi_{s}(x,y,t)\right |\\
      =&\left | \int dy\ H(y,t) S(y) \right |
\end{aligned}
\label{eoverlap} \end{equation}
and 
\begin{equation} 
\begin{aligned}
H(y,t)=&\int dEdE' \ \frac{\left | A(y_{0})\right |^{2}a^{*}(E')a(E)}{\hbar\sqrt{k(y,E)k(y,E')}}e^{-i(E-E')t/\hbar} \\
&\times e^{i(\varphi (E,y_{0})-\varphi (E',y_{0})+\int_{y_{0}}^{y}(k(y',E)-k(y',E'))dy')}\\
S(y)=&\ e^{-\frac{x_{0}^{2}}{4\sigma(y)^{2}}}
\end{aligned}
\end{equation}
\begin{figure*}
\includegraphics[width=17.2cm]{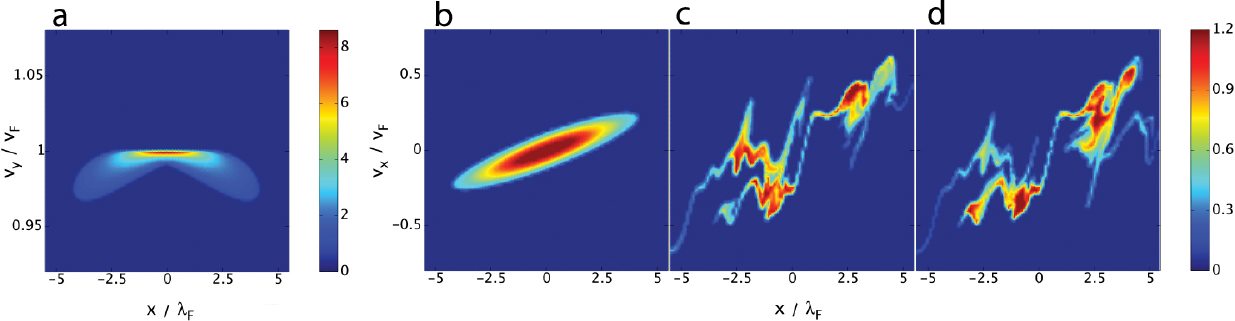}
\caption{(Color Online) Poincare Surface of Section at y=$y_{0}$+25$\lambda_{F}$. ({\bf a}) $\&$ ({\bf b}) are calculated in the absence of disorder, but disorder is present in ({\bf c}) $\&$ ({\bf d}). ({\bf c}) is plotted before the QPC is shifted and ({\bf d}) is when the QPC is shifted by $\lambda_{F}$ to the right. The color axis shows $\sqrt{P(x,v_{y})}$ in ({\bf a}) and $\sqrt{P(x,v_{x})}$ in the rest, all in unit of $1/\sqrt{v_{F}\lambda_{F}}$, where $v_F$ is the Fermi velocity and $v_x$ and $v_y$ are the velocity in the x and y direction respectively. More information about the method and this figure can be found in the supplementary material.}
\label{surface}\end{figure*}
H(y,t) is essentially a function needed for normalization and the integral can be estimated by considering only S(y). When $x_{0}$ is 0, $S(y)=1$ and normalization guarantees that $C_{o,s}(t)=1$. Initially, the wavepackets are centered around $y=y_{0}$ and we could choose an initial displacement $x_{0}\gg \sigma(y_{0})$ such that $S(y_0) \sim 0$ and $C_{o,s}(0) \sim 0$ . As time increases, the wavepacket will move away from the injection point and broaden. At typical experimental temperatures, the broadening is small compared with the distance it travels in y\cite{a4}. When the centers of the wavepackets reach a region far from the the injection point, $\sigma(y)$ around the new centers will grow to be much larger than $x_0$ and we have $S(y) \sim 1$, $C_{o,s}(t) \sim 1$. In other words, even though we start with two almost nonoverlapping(incoherent) initial wavepackets, the QPCs increase the coherent overlap as the wavepackets move away and this coherent overlap can reach unity in the far region. This overlap is of coherent nature and is different from the trivial spatial overlap one might expect. Spatial overlap is not enough to explain the experimentally observed stability due to the fluctuating phase in chaotic systems. However, our result shows that the overlap is large even if one takes into account the phases of the wavepackets and this coherent overlap can not be destroyed by disorders. This large coherent overlap only exists because the two different QPCs represent two different Hamiltonians. If propagated under the same Hamiltonian, the coherent overlap will always remain small.

In the experiment, both the finite size of the QPCs (making them not perfectly adiabatic) and the disorder can degrade the coherent overlap. We numerically estimate the coherent overlap under these conditions. The QPC's size is estimated from the Scanning Gate Microscopy data in \cite{a1} and the random potential has a correlation length of  $0.9\lambda_F$ and  standard deviation of $8\%E_F$, where $\lambda_F$ is the Fermi wavelength and $E_F=7.5meV$ is chosen to match that in the experiment. The random potential is generated to match both the sample mobility and the distance from donors to 2DEGs\cite{a19}. The numerical results show that $C_{o,s}(t)$= $85\%$ when the QPC is shifted by $\lambda_{F}$ as in the experiment. Thus the degradation of the coherent overlap at long range from the QPC is modest. 

Starting with two almost nonoverlapping(incoherent) initial wavepackets, and evolving separately under the influence of two different QPCs, their coherent overlap increases with time and distance, increasing fastest close to the QPCs. The coherent overlap eventually saturates to some constant value far from the QPCs. The two wavepackets now evolve effectively under the same Hamiltonian and their coherent overlap cannot be changed by the presence of disorder, for  example.  This is why we measure  25$\lambda_{F}$ downstream from the injection point, where the potential due to the QPC has died off. \textit{Given the large coherent overlap between the two wavepackets, we should expect the same set of branches far from the injection point even though the flow patterns look different close to the QPC, as shown in our quantum simulations in Fig.\ref{qflux}}.

Another kind of  disorder can reduce the overlap: backscattering from hard impurity scatterers. However, backscattering was suppressed in the original experiment due to the high purity of the samples used\cite{a1}.  
 

Reference\cite{a1} included both classical and quantum simulations and discussion. Does our explanation of the branch stability also apply to  classical simulations? Indeed it does, but the  proper classical initial conditions to represent the QPC are subtle and require care. A choice that closely resembles the quantum initial conditions is to use the Wigner quasiprobability distribution\cite{a7}, defined as
\begin{equation} 
P(\vec{x},\vec{p})=\frac{1}{\pi\hbar}\int_{-\infty}^{\infty}d\vec{s}\ e^{2i\vec{p}\cdot \vec{s}/\hbar}\Psi^{*}(\vec{x}+\vec{s})\Psi(\vec{x}-\vec{s})
\label{eWigner} \end{equation}
 The advantages are twofold: a) it produces the correct quantum spatial distribution $P(\vec{x})=\int P(\vec{x},\vec{p})d\vec{p}=|\Psi(\vec{x})|^{2}$ and momentum distribution $P(\vec{p})=\int P(\vec{x},\vec{p})d\vec{x}=|\Psi(\vec{p})|^{2}$; b) it properly accounts for the momentum uncertainty due to the confinement of QPC.  Keeping y fixed at $y_{0}$, applying (\ref{eWigner}) to (\ref{eigenchannel}) in x yields
\begin{equation} 
P(x,p_{x})=\frac{1}{\pi \sigma_{p_{x}} \sigma_{x}} e^{-\frac{p_{x}^2}{\sigma_{p_{x}}^{2}}-\frac{x^{2}}{\sigma_{x}^{2}}}
\label{eigenwi} \end{equation}
where $\sigma_{p_{x}}^{2}=m\hbar\omega(y_{0})$ and $\sigma_{x}^{2}=\hbar/m\omega(y_{0}) $

When $\omega(y)$ changes sufficiently slowly compared to the motion in y,  (\ref{eigenwi}) holds approximately true for any $y>y_{0}$, which implies that momentum distributions are highly correlated at any position no matter which QPC the electron originates from. The only difference is the overall probability of arriving at that point. This already  hints as to why branches remain at the same positions with a modified strength.

As in the quantum case, we  use a realistic QPC potential,  and weak random potentials in the open regions of the 2DEGs. We sample according to (\ref{eigenwi}),  with the keeping energy fixed at $E_{F}$ by eliminating trajectories with larger energy in the Wigner distribution, and boosting those with less in $p_y$.  These details may be omitted and do not change the conclusions about branch populations and overlap. We propagate the electrons classically.  We show  Poincare' surface of section plots\cite{a6,a19} in Fig.\ref{surface}.  In the absence of disorder(Fig.\ref{surface}a and b), the adiabaticity of the QPC ensures that when the electrons emerge, most energy is transferred from the x (transverse) direction to the y (longitudinal) direction, which is also expected in the quantum case. The results when disorder is present is shown in Fig.\ref{surface}c$\&d$. As can be seen, very similar regions in phase space are occupied, with different relative strengths, when the QPC is shifted. To quantify the overlap in phase space, we define the correlation to be
\begin{equation} 
C(P_{o},P_{s})=\int dxdp_{x} \ \sqrt{P_{o}(x,p_{x})P_{s}(x,p_{x})}
\end{equation}
where $P_{o}$ corresponds to the distribution in the original QPC and  $P_{s}$ the shifted one. Twenty five wavelengths away from the injection point, it is measured that $C(P_{o},P_{s})$=79$\%$, which is comparable to our quantum result. 

\begin{figure}
\includegraphics[width=8cm]{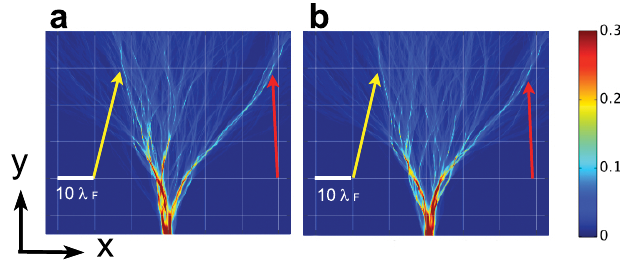}
\caption{(Color Online) Classical simulations of the total flux in the y direction that passes through a given point. ({\bf a}) shows the case where the QPC is not shifted and ({\bf b}) shows that when the QPC is shifted by $\lambda_{F}$ to the right. The white reference grid is at the same location in both images and the plots start at y=$y_{0}+15\lambda_{F}$. In both figures, the branches labeled by the red and yellow arrows are clearly visible.  More information about this figure can be found in the supplementary material. }
\label{cflux}\end{figure}

The classical approach to branching is based on  caustics which develop in coordinate space due to focussing effects, and stable regions in phase space\cite{a8,a5,a16} which persist some distance away from the injection point. (Eventually, stable regions, which form by chance so to speak in the random potential,  are also subject to destruction further on in the random potential). Each branch corresponds to a localized region in phase space with its strength determined by the electron density in those regions. After shifting the QPC, similar regions in phase space are occupied with only a changed relative density, which means in coordinate space that the same  branches are occupied with a different strength. This explains the observation in the experiment\cite{a1}. In Fig.\ref{cflux}, we present our simulations of the total classical flux that passes through a given point, which confirms that classical trajectories can indeed reproduce the observed stability. It is worth noting that  the same stable regions could in principle be populated from both QPCs, causing some similarity of branch appearance, but this will not be a generic effect for all random potentials and QPC shifts\cite{a17}.

\begin{figure}
\includegraphics[width=6cm]{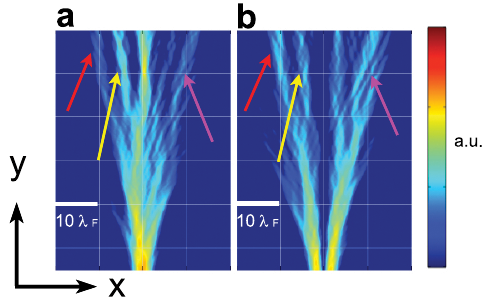}
\caption{(Color Online) Quantum simulations of total flux in the y direction that passes through a given point when the second mode of the QPC is open. ({\bf a}) corresponds to the first mode of the QPC while ({\bf b}) shows contribution from the second mode alone. The starting point and length scale are the same as in Fig.\ref{qflux}, but the flux strength is presented in log scale instead. In both figures, the branches pointed to by the red, yellow and purple arrows are clearly visible. }
\label{second}\end{figure}

One advantage of a classical interpretation is that it can provide intuition in cases where the quantum dynamics is less intuitive. One example would be to consider what happens when the second mode of QPC is open. According to reference \cite{a4}, we need to independently propagate two wavepackets where one corresponds to the first mode and the other corresponds to the second mode. Their contributions to the flux are then added up incoherently to produce the experimental measurements. Since the contribution from the first mode is added incoherently, it is no surprise that the same set of branches recurs when both modes are open. However, it is interesting to ask what happens if one looks at the contribution from each mode alone. Quantum mechanically, the first and second mode are orthogonal to each other, and, therefore have zero overlap at all time since the Hamiltonian is the same. However, the classical phase space regions corresponding to the second mode alone would still overlap more or less with that due to the first mode\cite{a18}. As a result, the second mode alone should still produce some similar branches that appear in the first mode with a different strength. In order to see this effect, we take the logarithm of the quantum flux due to each mode alone and present the results in Fig.\ref{second}. As we can see, some of the strongest branches are clearly preserved, which verifies our prediction.

In conclusion, we have successfully explained the stability of branched flow against large changes in initial conditions using both quantum and classical simulations, which agree on the fact of the stability of branches against shifts of the QPC injection point. This resolves a puzzle raised by a recent experiment\cite{a1} and shows the role of the QPC in enhancing the stability of branched flow in 2DEGs. Our classical interpretation predicts a further stability of the branched flow that can not be readily inferred form the experiment. The interpretations in this paper can provide useful insights into future applications in the coherent control of electron flow, branch management and probing local random potential.

We acknowledge support from Department of Energy under DE-FG02-08ER46513.


\begin{thebibliography}{99}

 \bibitem{a2} 
M.A. Topinka,  B. J. LeRoy, S. E. J. Shaw,  E. J. Heller,  R. M. Westervelt, K. D. Maranowski and A. C. Gossard, Science {\bf 289}, 2323 (2000).
 \bibitem{a3}  
M. A. Topinka, B. J. LeRoy, R. M. Westervelt, S. E. J. Shaw, R. Fleischmann, E. J. Heller, K. D. Maranowski and A. C. Gossard, Nature  {\bf 410}, 183-186 (2001).
 \bibitem{a12} 
R. Hohmann, U. Kuhl, H.-J. Stockmann, L. Kaplan and E. J. Heller, Phys. Rev. Lett {\bf 104}, 093901 (2010).
\bibitem{a10}
M. A. Wolfson and S. Tomsovic,  J. Acoust. Soc. Am. {\bf 109}, 2693 (2001).
\bibitem{a14}
K. E. Aidala, R. E. Parrott, T. Kramer, E. J. Heller, R. M. Westervelt, M. P. Hanson and A. C. Gossard, Nat. Phys. {\bf 3}, 464 (2007).
 \bibitem{a13} 
D. Maryenko, F. Ospald, K. V. Klitzing, J. H. Smet,  J. J. Metzger, R. Fleischmann, T. Geisel and V. Umansky, Phys. Rev. B {\bf 85}, 195329 (2012).
\bibitem{a9}
E. J. Heller, L. Kaplan, and A. Dahlen,  J. Geophys.Res.  {\bf 113}, C09023(2008).
 \bibitem{a1}
M. P. Jura, M. A. Topinka, L. Urban, A. Yazdani, H. Shtrikman, L. N. Pfeiffer, K. W. West and D. Goldhaber-Gordon, Nat. Phys. {\bf 3}, 841-845 (2007).
 \bibitem{a4}  
E. J. Heller, K. E. Aidala, B. J. LeRoy, A. C. Bleszynski, A. Kalben, R. M. Westervelt, K. D. Maranowski and A. C. Gossard, Nano Lett.  {\bf 5}, 1285 (2005).
\bibitem{a6}
S. E. J. Shaw's Thesis, Harvard University, available at http://www.physics.harvard.edu/Thesespdfs/sshaw.pdf (2002).
 \bibitem{a15} 
Adel Abbout, Gabriel Lemarie and Jean-Louis Pichard, Phys. Rev. Lett {\bf 106}, 156810 (2011).
\bibitem{a7}
E. J. Heller,  J. Chem. Phys. {\bf 65}, 1289 (1976).
 \bibitem{a8} 
L. Kaplan, Phys. Rev. Lett {\bf 89}, 184103 (2002).
 \bibitem{a5} 
E. J. Heller and S. E. J. Shaw, International Journal of Modern Physics B  {\bf 17}, 3977 (2003).
 \bibitem{a16} 
J. J. Metzger, R. Fleischmann and T. Geisel, Phys. Rev. Lett {\bf 105}, 020601 (2010).
\bibitem{a17} 
If we have a stable region in phase space that happens to be cut into halves by the shifting of the QPC, it will produce close branches with similar strength even with zero overlap at all times. However, there are two reasons why this will only have marginal effect in producing the experimentally observed stability. First, the probability of cutting a stable regions in halves such that both regions have appreciable strength is negligible. Therefore, it is not reproducible by experiment. Secondly and most importantly, any classical stable regions will decay with distance. Thus, any close branches, if exist, will separate eventually and contribute only marginally to the experimental stability. 
 \bibitem{a18} 
This doesn't depend on the choice of the Wigner quasiprobability distribution. It is simply due to the fact that both modes have to live in the same region in coordinate space and maintain considerable amount of momentum uncertainty due to the confinement.
 \bibitem{a19} 
See the supplementary material provided for more information.


\end{thebibliography}
\end{document}


\title{Supplemental Material for ``Stability of Branched Flow from a Quantum Point Contact''}

\author{Bo Liu} \affiliation{Department of Physics, Harvard University, Cambridge, Massachusetts 02138, USA}
\author{Eric J. Heller} \affiliation{Department of Physics, Harvard University, Cambridge, Massachusetts 02138, USA}
\affiliation{Department of Chemistry and Chemical Biology, Harvard University, Cambridge, Massachusetts 02138, USA}

\date{\today}

\maketitle
In this supplementary material, we present the theoretical background of our approach and the details of the numerical methods used in the paper. 
\\
\\
\section*{Quanutm Simulation Methods }
The method used to calculate the quantum flow pattern is mainly developed in reference \cite{a1} and \cite{a2} and most of the materials in this section can be found there. 

In the experiments\cite{a3,a4,a6}, the electron flow pattern is measured through the change in conductance as a function of the position of the scanning tip. Thus the starting point for the theory is the Landauer formula at finite temperature, which is given by 
\begin{equation} 
\begin{aligned}
G_T=\frac{2e^2}{h}\int \frac{-\partial f_T(E)}{\partial E} Tr[\mathbf{t}(E)\mathbf{t}(E)^{+}]dE
\end{aligned}
\label{H} \end{equation}
where $G_T$ is the conductance at temperature T, $f_T(E)$ is the corresponding Fermi distribution and $\mathbf{t}(E)$ is the transmission matrix. In the presence of a QPC, there will only be a discrete number of modes open for transmission at any given energy E. Since we only care about the trace in (\ref{H}), we can choose to work with the eigenmode of the QPC and add up the contribution from each mode separately, which is the underlying idea for the so-called "thermal wavepacket" approach developed in \cite{a1}.

When only one mode is open in the QPC, equation (\ref{H}) reduces to 
\begin{equation} 
\begin{aligned}
G_T=\frac{2e^2}{h}\int \frac{-\partial f_T(E)}{\partial E} \tau_1(E)dE
\end{aligned}
\label{H1} \end{equation}
where $\tau_1(E)$ is the only nonzero eigenvalue of $\mathbf{t}(E)\mathbf{t}(E)^{+}$.

In this case, we can create an initial wavepacket of the following form
\begin{equation} 
\begin{aligned}
\Psi_{1,T}(x,y,0)&=\int dE \ e^{i\varphi (E,y_{0})}a_T(E)\Psi _{1}(x,y,E)
\end{aligned}
\label{iwave} \end{equation}
where $\Psi _{1}(x,y,E)$ is the scattering eigenstate with transmission probability $\tau_1(E)$, $\varphi (E,y_{0})$ is chosen such that it is a compact wavepacket centered at $y_0$ and $a_T(E)$ represents the energy distribution.

It is shown in \cite{a1} that the total transmitted flux is given by the following formula
\begin{equation} 
\begin{aligned}
F_T=\frac{4\pi e^2 \hbar}{m}\int \left | a_T(E) \right |^2 \tau_1(E)dE
\end{aligned}
\label{flux} \end{equation}
If we choose $a_T(E)=\hbar^{-1}\sqrt{-\frac{m}{2\pi}\frac{\partial f_T(E)}{\partial E}}$, we would get $F_T=G_T$, which shows that one can get all the information about the experimental conductance map by propagating this wavepacket through the scattering region.

For any given point, the time dependence of the flux can be calculated by applying the following flux operator to the wavepacket\cite{a2}:
\begin{equation} 
\begin{aligned}
\hat{f}_s=\frac{1}{2}[\hat{n}\cdot\hat{v}\delta(\vec{r}-\vec{r}_s)+\delta(\vec{r}-\vec{r}_s)\hat{n}\cdot\hat{v}]
\end{aligned}
\label{flux} \end{equation}
where $\hat{n}$ is the direction of interest, which is chosen to be the y direction in the simulations. $\vec{r}_s$ is the point of interest and $\hat{v}=-\frac{i\hbar}{m}\triangledown $ is the velocity operator. For our quantum simulations in Fig.1, we report the total flux through a given point, which is $\int \Psi_{1,T}^*(x,y,t)\hat{f}_s\Psi_{1,T}(x,y,t)dtdxdy$.

When the second transmitting mode of the QPC is also open, we need to propagate a second wavepacket separately to include its contribution. The total quantum flux will be an incoherent sum of the flux from both wavepackets. This is so because  equation (\ref{H}) only involves the trace of the scattering matrix. Just as what we do for the first mode, we can prepare a second initial wavepacket of the following form:
\begin{equation} 
\begin{aligned}
\Psi_{2,T}(x,y,0)&=\int dE \ e^{i\varphi' (E,y_{0})}a_T(E)\Psi_{2}(x,y,E)
\end{aligned}
\label{iwave} \end{equation}
where $\Psi_{2}(x,y,E)$ is the scattering eigenstate corresponding to the second eigenmode of the QPC , $\varphi '(E,y_{0})$ is chosen such that it is a compact wavepacket centered at $y_0$ and $a_T(E)=\hbar^{-1}\sqrt{-\frac{m}{2\pi}\frac{\partial f_T(E)}{\partial E}}$.  To reproduce the experimental results, one just needs to add up the contributions from $\Psi_{1,T}(x,y,0)$ and $\Psi_{2,T}(x,y,0)$ incoherently. However, in our case, we are interested in the contributions from each mode alone, so we present the results separately in Fig.4.
\\

\section*{Random Potential Generation}
There are two major kinds of disorders in the experimental system\cite{a5}. The first one is from the charged donor atoms in the donor layer, whose distance to the 2DEGs can be controlled experimentally. This distance is related to the correlation length of the random potential. The second one is from the impurity atoms, which can lie very close to the 2DEGs, resulting in a sharp potential that strongly backscatters. It is established both theoretically\cite{a1,a5} and experimentally\cite{a6} that the random potential from the donors is mainly responsible for branching while the impurity potential is responsible for fringing, which is another striking observation of the experiment\cite{a3,a4}.  For the experimental results\cite{a6} we are interested in, fringing is significantly suppressed by using very high mobility samples\cite{a6}. Thus, we shall focus only on the random potential due to donor atoms. 

One simple model that takes into account screening yields the following result\cite{a2} for the potential due to one charged donor atom:
\begin{equation} 
\begin{aligned}
V(r)=-q\frac{\hbar^2}{2m} \frac{d}{(r^2+d^2)^{3/2}}
\end{aligned}
\label{flux} \end{equation}
where q is the charge on the donor atom, d is the donor to 2DEGs distance and r is the distance measured in-plane.

Given an estimate of donor density in the experiment, one can then generate a random potential using the above model. However, in order to match the mobility of the experimental samples, one needs to adjust the standard deviation of the strength of the random potential. As estimated by both us and the experimental group in their supplementary material\cite{a6}, the resulting random potential has a standard deviation of $8\% E_F$. 

To characterize the legnth scale in the random potential, we define a correlation length through the autocorrelation function of the potential. Given a random potential $V(\vec{r})$, the autocorrelation function is defined to be $<V(\vec{r}'-\vec{r})V(\vec{r}')>$, where the bracket means sample or ensemble average over $\vec{r}'$.  It is approximately true that the autocorrelation function has the following scaling relation:

\begin{equation} 
<V(\vec{r}'-\vec{r})V(\vec{r}')> \sim e^{-\frac{r^2}{l_c^2}} 
\end{equation}
 where $l_c$ is defined to be the correlation length.  For our sample, the correlation length is estimated to be $l_c=0.9\lambda_F$.
\\

\section*{Classical Simulation Methods }
For the classical simulations, we use Monte Carlo method to generate $10^7$ classical particles with the following initial distribution in position and momentum:
\begin{equation} 
P(x,p_{x})=\frac{1}{\pi \sigma_{p_{x}} \sigma_{x}} e^{-\frac{p_{x}^2}{\sigma_{p_{x}}^{2}}-\frac{x^{2}}{\sigma_{x}^{2}}}
\label{eigenwi} \end{equation}
where $\sigma_{p_{x}}^{2}=m\hbar\omega(y_{0})$ and $\sigma_{x}^{2}=\hbar/m\omega(y_{0}) $.  The energy of each particle is chosen to be fixed at $E_F$. This is guaranteed by eliminating trajectories with larger energy and boosting those with less in $p_y$.The time evolution of those particles is governed by the classical equation of motion and the results are presented by the method of Poincare's surface of section\cite{a2} at fixed y in Fig.2 in the paper. Essentially, the idea behind Poincare's surface of section is to reduce the dimension of phase space to two by fixing y. Each classical particle can be represented in the four dimensional phase space by $(x,y,p_x,p_y)$. Since we are dealing with classical particles of fixed energy, the constraint imposed by the conservation of energy reduces the phase space structure to three dimensions. The two dimension plots shown in Fig.2 are done by further fixing y, which decreases the degree of freedom by one. If one is interested in recovering the full three dimensional structure in phase space, one can plot many two-dimensional plots at different y and combine them. For our purpse, we are only interested in the saturated overlap produced by the QPC. Therefore, we only show plots at a y that is sufficiently far away from the QPC, where the overlap has saturated to a constant value. 

For the classical flux simulations shown in Fig.3, we use the same Monte Carlo methods to sample the distribution given by (\ref{eigenwi}) and calculate the total flux through a given point by counting the total number of classical particles that pass through a small neighborhood of that point with weights determined by their momenta in the y direction.\\
\\
\\